\documentclass[pdftex,letterpaper,rmp,twocolumn,groupedaddress,floatfix]{revtex4}
\pdfoutput=1
\bibpunct{(}{)}{,}{n}{}{;}
\usepackage[utf8]{inputenc}
\usepackage{graphicx}
\usepackage{amsfonts}
\usepackage{amsmath}
\usepackage{amssymb}
\usepackage{color}
\definecolor{darkgray}{rgb}{0.25,0.25,0.25}
\definecolor{darkred}{rgb}{0.89,0.10,0.11}
\definecolor{darkblue}{rgb}{0.12,0.39,0.62}
\usepackage{url}
\usepackage[pdftex,breaklinks=true,colorlinks=true,citecolor=black,linkcolor=black,menucolor=black,urlcolor=darkblue,pdfborder={1 0 0}]{hyperref}
\hypersetup{pdftitle={The map equation},pdfauthor={Martin Rosvall, Daniel Axelsson, and Carl Bergstrom, 2009}}
\begin{document}

\title{The map equation}
\author{M. Rosvall}
\email{martin.rosvall@physics.umu.se}
\homepage{http://www.tp.umu.se/~rosvall/}
\affiliation{Department of Physics, Umeå University, SE-901 87 Umeå, Sweden}
\author{D. Axelsson}
\affiliation{Department of Physics, Umeå University, SE-901 87 Umeå, Sweden}
\author{C. T. Bergstrom}
\affiliation{Department of Biology, University of Washington, Seattle, WA 98195-1800}
\affiliation{Santa Fe Institute, 1399 Hyde Park Rd., Santa Fe, NM 87501}
\date{\today}

\begin{abstract}
	
Many real-world networks are so large that we must simplify their structure before we can extract useful information about the systems they represent. As the tools for doing these simplifications proliferate within the network literature, researchers would benefit from some guidelines about which of the so-called community detection algorithms are most appropriate for the structures they are studying and the questions they are asking. Here we show that different methods highlight different aspects of a network's structure and that the the sort of information that we seek to extract about the system must guide us in our decision. For example, many community detection algorithms, including the popular modularity maximization approach, infer module assignments from an underlying model of the network formation process. However, we are not always as interested in how a system's network structure was formed, as we are in how a network's extant structure influences the system's behavior. To see how structure influences current behavior, we will recognize that links in a network induce movement across the network and result in system-wide interdependence. In doing so, we explicitly acknowledge that most networks carry flow. To highlight and simplify the network structure with respect to this flow, we use the map equation. We present an intuitive derivation of this flow-based and information-theoretic method and provide an interactive on-line application that anyone can use to explore the mechanics of the map equation. The differences between the map equation and the modularity maximization approach are not merely conceptual. Because the map equation attends to patterns of flow on the network and the modularity maximization approach does not, the two methods can yield dramatically different results for some network structures. To illustrate this and build our understanding of each method, we partition several sample networks. We also describe an algorithm and provide source code to efficiently decompose large weighted and directed networks based on the map equation.  
\end{abstract}

\maketitle

\section{Introduction}

Networks are useful constructs to schematize the organization of interactions in social and biological systems. Networks are particularly valuable for characterizing \emph{interdependent} interactions, where the interaction between components A and B influences the interaction between components B and C, and so on. For most such integrated systems, it is a flow of some entity --- passengers traveling among airports, money transferred among banks, gossip exchanged among friends, signals transmitted in the brain --- that connects a system's components and generates their interdependence. Network structures constrain these flows. Therefore, understanding the behavior of integrated systems at the macro-level is not possible without comprehending the network structure with respect to the flow, the \emph{dynamics on} the network.

One major drawback of networks is that, for visualization purposes, they can only depict small systems. Real-world networks are often so large that they must be represented by coarse-grained descriptions. Deriving appropriate coarse-grain descriptions is the basic objective of community detection \cite{girvan_newman,palla,guimerahierarchy,fortunato}. But before we decompose the nodes and links into modules that represent the network, we must first decide what we mean by ``appropriate." That is, we must decide which aspects of the system should be highlighted in our coarse-graining. 

 If we are concerned with the process that {\em generated} the network in the first place, we should use methods based on some underlying stochastic model of network formation. To study the formation process, we can, for example, use modularity \cite{girvan}, mixture models at two \cite{newman2007mma} or more \cite{clauset2008hsa} levels, Bayesian inference \cite{hofman2008ban}, or our cluster-based compression approach \cite{RosvallBergstrom07} to resolve community structure in undirected and unweighted networks. 
If instead we want to infer system behavior from network structure, we should focus on how the structure of the extant network constrains the dynamics that can occur on that network. To capture how local interactions induce a system-wide flow that connects the system, we need to simplify and highlight the underlying network structure with respect to how the links drive this flow across the network. For example, both Markov processes on networks and spectral methods can capture this notion \cite{ziv,donath,kasper,delvenne2008sgc}. In this paper, we present a detailed description of the flow-based and information-theoretic method known as the map equation \cite{RosvallBergstrom08}. For a given network partition, the map equation specifies the theoretical limit of how concisely we can describe the trajectory of a random walker on the network. With the random walker as a proxy for real flow, minimizing the map equation over all possible network partitions reveals important aspects of network structure with respect to the dynamics on the network. To illustrate and further explain how the map equation operates, we compare its action with the topological method modularity maximization \cite{girvan_newman}. Because the two methods can yield different results for some network structures, it is illuminating to understand when and why they differ.

\section{Mapping flow}

There is a duality between the problem of compressing a data set, and the problem of detecting and extracting significant patterns or structures within those data. This general duality is explored in the branch of statistics known as MDL, or minimum description length statistics \cite{rissanen1978,grunwald}. We can apply these principles to the problem at hand: finding the structures within a network that are significant with respect to how information or resources flow through that network. 

To exploit the inference-compression duality for dynamics on networks, we envision a communication process in which a sender wants to communicate to a receiver about movement on a network. That is, we represent the data that we are interested in --- the trace of the flow on the network --- with a compressed message. 
This takes us to the heart of information theory, and we can employ Shannon's source coding theorems to find the limits on how far we can compress the data \cite{shannon48}. For some applications, we may have data on the actual trajectories of goods, funds, information, or services as they travel through the network, and we could work with the trajectories directly. More often, however, we will only have a characterization of the network structure along which these objects can move, in which case we can do no better than to approximate likely trajectories as random walks guided by the directed and weighted links of the network. This is the approach that we take with the map equation.

In order to effectively and concisely describe where on the network a random walker is, an effective encoding of position will necessarily exploit the regularities in patterns of movement on that network. If we can find an optimal code for describing places traced by a path on a network, we have also solved the dual problem of finding the important structural features of that network. Therefore, we look for a way to assign codewords to nodes that is efficient with respect to the dynamics on the network. 

\begin{figure*}[thbp]
\centering
\includegraphics[width=\textwidth]{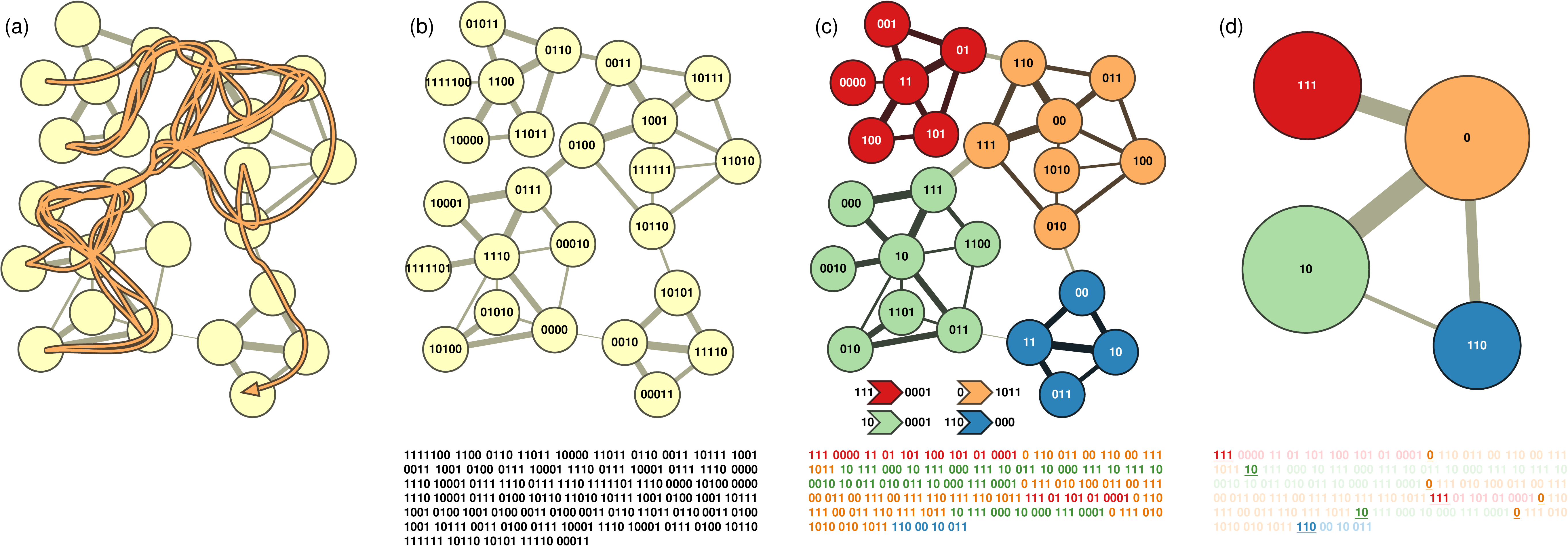}
\caption{\label{fig1}
Detecting regularities in patterns of movement on a network (derived from \cite{RosvallBergstrom08} and also available in a dynamic version \cite{flashapplet}). (a) 
We want to effectively and concisely describe the trace of a random walker on a network. The orange line shows one sample trajectory. For an optimally efficient one-level description, we can use the codewords of the Huffman codebook depicted in (b). The 314 bits shown under the network describes the sample trajectory in (a), starting with $1111100$ for the first node on the walk in the upper left corner, $1100$ for the second node, etc., and ending with $00011$ for the last node on the walk in the lower right corner. (c) A two-level description of the random walk, in which an index codebook is used to switch between module codebooks, yields on average a 32\% shorter description for this network. The codes of the index codebook for switching module codebooks and the codes used to indicate an exit from each module are shown to the left and the right of the arrows under the network, respectively. Using this code, we capitalize on structures with long persistence times, and we can use fewer bits than we could do with a one-level description. For the walk in (a), we only need the 243 bits shown under the the network in (c). The first three bits $111$ indicate that the walk begins in the red module, the code $0000$ specifies the first node on the walk, and so forth. (d) Reporting only the module names, and not the locations within the modules, provides an efficient coarse-graining of the network.}
\end{figure*}

A straightforward method of assigning codewords to nodes is to use a Huffman code \cite{huffman}. Huffman codes are optimally efficient for symbol-by-symbol encoding and save space by assigning short codewords to common events or objects, and long codewords to rare ones, just as Morse code uses short codes for common letters and longer codes for rare ones. Figure \ref{fig1}(b) shows a prefix-free Huffman coding for a sample network. It corresponds to a lookup table for coding and decoding nodes on the network, a \emph{codebook} that connects nodes with codewords. In this codebook, each Huffman codeword specifies a particular node, and the codeword lengths are derived from the ergodic node visit frequencies of a random walk (the average node visit frequencies of an infinite-length random walk). Because the code is prefix-free, that is, no codeword is a prefix of any other codeword, codewords can be sent concatenated without punctuation and still be unambiguously decoded by the receiver. With the Huffman code pictured in Fig.~\ref{fig1}(b), we are able to describe the nodes traced by the specific 71-step walk in 314 bits. If we instead had chosen a uniform code, in which all codewords are of equal length, each codeword would be  $\lceil \log{25} \rceil = 5$ bits long (logarithm taken in base 2), and $71\cdot5=355$ bits would have been required to describe the walk.

This Huffman code is optimal for sending a one-time transmission describing the location of a random walker at one particular instant in time. Moreover, it is optimal for describing a list of locations of the random walker at arbitrary (and sufficiently distant) times. However, if we wish to list the locations visited by our random walker in a sequence of successive steps, we can do better. Sequences of successive steps are of critical importance to us; after all, this is flow.

Many real-world networks are structured into a set of regions such that once the random walker enters a region, it tends to stay there for a long time, and movements between regions are relatively rare.  As we design a code to enumerate a succession of locations visited, we can take advantage of this regional structure. We can take a region with a long persistence time and give it its own separate codebook. So long as we are content to reuse codewords in other regional codebooks, the codewords used to name the locations in any single region will be shorter than those in the global Huffman code example above, because there are fewer locations to be specified. We call these regions ``modules'' and their codebooks ``module codebooks.''  However, with multiple module codebooks, each of which re-uses a similar set of codewords, the sender must also specify which module codebook should be used. That is, every time a path enters a new module, both sender and receiver must simultaneously switch to the correct module codebook or the message will be nonsense. This is implemented by using one extra codebook, an index codebook, with codewords that specify which of the module codebooks is to be used. The coding procedure is then as follows. The index codebook specifies a module codebook, and the module codebook specifies a succession of nodes within that module. When the random walker leaves the module, we need to return to the index codebook. To indicate this, instead of sending another node name from the module codebook, we send the ``èxit command'' from the module codebook. The codeword lengths in the index codebook are derived from the relative rates at which a random walker enters each module, while the codeword lengths for each module codebook are derived from the relative rates at which a random walker visits each node in the module or exits the module.

Here emerges the duality between coding a data stream and finding regularities in the structure that generates that stream. Using multiple codebooks, we transform the problem of minimizing the description length of places traced by a path into the problem of how we should best partition the network with respect to flow. How many modules should we use, and which nodes should be assigned to which modules to minimize the map equation? Figure \ref{fig1}(c) illustrates a two-level description that capitalizes on structures with long persistence time and encodes the walk in panel (a) more efficiently than the one-level description in panel (b). We have implemented a dynamic visualization and made it available for anyone to explore the inference-compression duality and the mechanics of the map equation (\url{http://www.tp.umu.se/~rosvall/livemod/mapequation/}).

\begin{figure*}[tbp]
\centering
\includegraphics[width=0.85\textwidth]{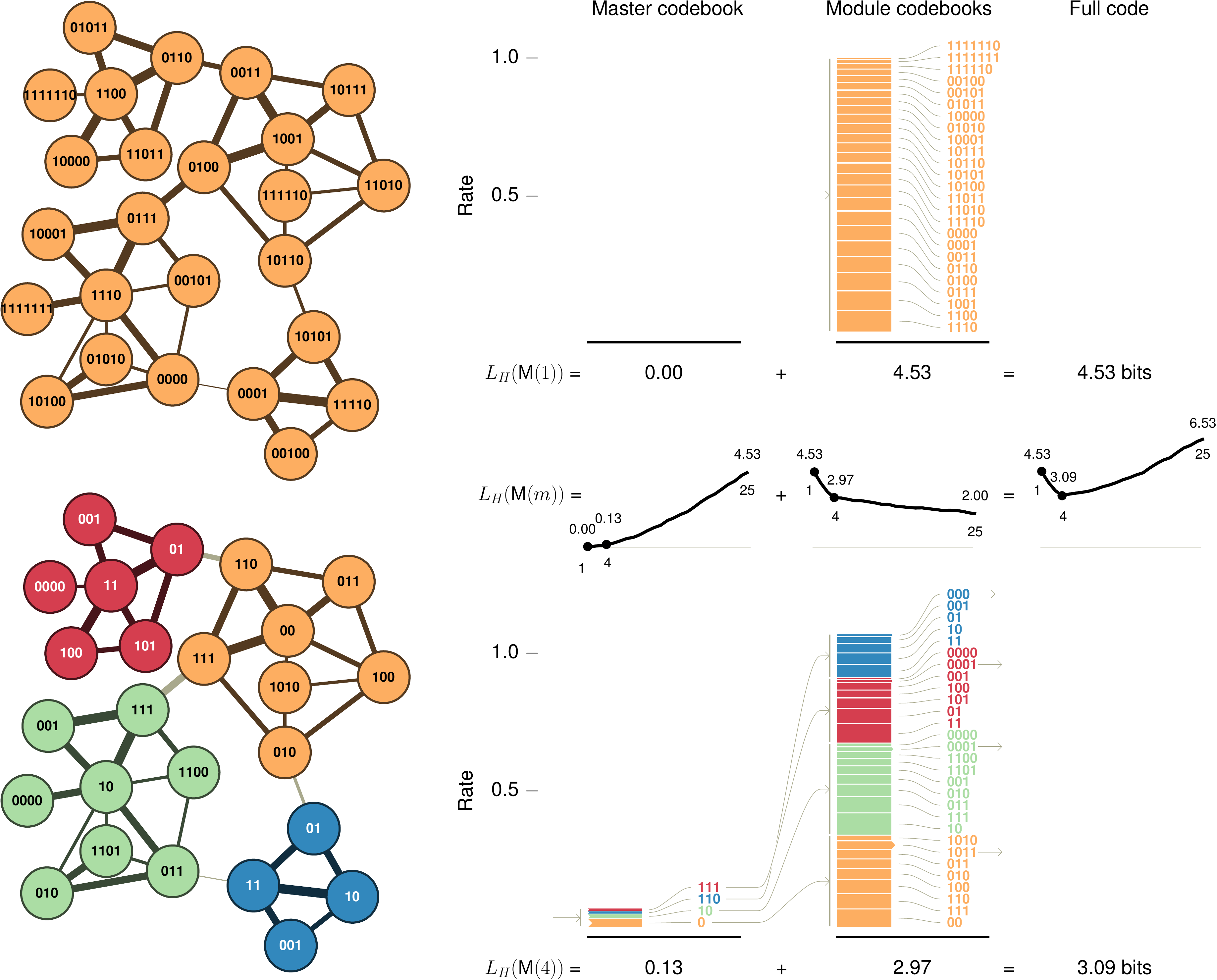}
\caption{\label{codebooks}
The duality between finding community structure in networks and minimizing the description length of a random walk across the network (available online as a Flash applet \cite{flashapplet}). With a single module codebook (top) we cannot capture the higher order structure of the network, and the per-step description length of the random walker's movements is on average 4.53 bits. The sparklines (middle) show how the description length for module transitions, within-module movements, and the sum of the two depend on the number of modules $m$ varying from 1 to 25 (with explicit values for the one- and four-module partitions included in this figure and the 25-module partition with one module for each node). The description length for module transitions increases monotonically, and the description length for within-module movements decreases monotonically with the number of modules. The sum of the two, the total description length $L_H(\mathsf{M}(m))$, with subscript $H$ for Huffman to distinguish this length from the Shannon limit used in the actual map equation, takes a minimum at four modules.
With an index codebook and four module codebooks (bottom), we can compress the description length down to an average of 3.09 bits per step. For this optimal partition of the network, the four module codebooks are associated with smaller sets of nodes and shorter codewords, but still with long persistence time and few between-module movements. This compensates for the extra cost of switching modules and accessing the index codebook. \\
To illustrate the duality between module detection and coding, the code structure corresponding to the current partition is shown by the stacked boxes on the right. The height of each box corresponds to the per-step rate of codeword use.
The left stack represents the index codebook associated with movements between modules. The height of each box is equal to the exit probability $q_{i \curvearrowright }$ of the corresponding module $i$. Boxes are ordered according to their heights. The codewords naming the modules are the Huffman codes calculated from the probabilities $q_{i \curvearrowright } / q_{\curvearrowright }$, where $q_{\curvearrowright }=\sum_1^m q_{i \curvearrowright }$ is the total height of the left stack. The length of the codeword naming module $i$ is approximately $-\log{q_{i \curvearrowright } / q_{\curvearrowright }}$, the Shannon limit in the map equation. In the figure, the per-step description length of the random walker's movements between modules is the sum of the length of the codewords weighted by their use. This length is bounded below by the limit $-\sum_1^m q_{i \curvearrowright }\log{q_{i \curvearrowright } / q_{\curvearrowright }}$ that we use in the map equation.\\
The right stack represents the module codebooks associated with movements within modules. The height of each box in module $i$ is equal to the ergodic node visit probabilities $p_{\alpha \in i}$ or the exit probability $q_{i \curvearrowright }$ marked by the blocks with arrows. The boxes corresponding to the same module are collected together and ordered according to their weight; in turn, the modules are ordered according to their total weights $p_{\circlearrowright}^i = q_{i \curvearrowright } + \sum_{\alpha \in i} p_\alpha$. The codewords naming the nodes and exits (the arrows on the box and after the codeword distinguish exits from nodes and illustrate that the index codebook will be accessed next) in each module $i$ are the Huffman codes calculated from the probabilities $p_{\alpha \in i}/p_{\circlearrowright}^i$ (nodes) and $q_{i \curvearrowright }/p_{\circlearrowright}^i$ (exits). The length of codewords naming nodes $\alpha \in i$ and exit from module $i$ are approximately $-\log(p_{\alpha \in i} /p_{\circlearrowright}^i)$ (nodes) and $-\log(q_{i \curvearrowright }/p_{\circlearrowright}^i)$ (exits). In the figure, the per-step description length of the random walker's movements within modules is the sum of the length of the codewords weighted by their use. This length is bounded below by the limit $-\sum_1^m \left[q_{i \curvearrowright }\log(q_{i \curvearrowright } / p_{\circlearrowright}^i) + \sum_{\alpha \in i}p_\alpha\log(p_\alpha / p_{\circlearrowright}^i)\right]$.}
\end{figure*}

Figure \ref{codebooks} visualizes the use of one or multiple codebooks for the network in Fig.~\ref{fig1}. The sparklines show how the description length associated with between-module movements increases with the number of modules and more frequent use of the index codebook. Contrarily, the description length associated with within-module movements decreases with the number of modules and with the use of smaller module codebooks. The sum of the two, the full description length, takes a minimum at four modules. We use stacked boxes to illustrate the rates at which a random walker visits nodes and enters and exits modules. The codewords to the right of the boxes are derived from the within-module relative rates and within-index relative rates, respectively. Both relative rates and codewords change from the one-codebook solution with all nodes in one module, to the optimal solution, with an index codebook and four module codebooks with nodes assigned to four modules (see online dynamic visualization \cite{flashapplet}). 

\section{The map equation}

We have described the Huffman coding process in detail in order to make it clear how the coding structure works. But of course the aim of community detection is not to encode a particular path through a network. In community detection, we simply want to find the modular structure of the network with respect to flow and our approach is to exploit the inference-compression duality to do so. In fact, we do not even need to devise an optimal code for a given partition to estimate  how efficient that optimal code would be. This is the whole point of the map equation. It tells us how efficient the optimal code would be for any given partition, without actually devising that code. That is, it tells us the theoretical limit of how concisely we can specify a network path using a given partition structure. To find an optimal partition of the network, it is sufficient to calculate this theoretical limit for different partitions of the network and pick the one that gives the shortest description length.

For a module partition $\mathsf{M}$ of $n$ nodes $\alpha=1,2,\ldots,n$ into $m$ modules $i=1,2,\ldots,m$, we define this lower bound on code length to be $L(\mathsf{M})$. To calculate $L$ for an arbitrary partition, we first invoke Shannon's source coding theorem \cite{shannon48}, which implies that when you use $n$ codewords to describe the $n$ states of a random variable $X$ that occur with frequencies $p_i$, the average length of a codeword can be no less than the entropy of the random variable $X$ itself: $H(X) = -\sum_{1}^{n} p_i \log(p_i)$ (we measure code lengths in bits and take the logarithm in base 2). This provides us with a lower bound on the average length of codewords for each codebook. 
To calculate the average length of the code describing a step of the random walk, we need only to weight the average length of codewords from the index codebook and the module codebooks by their rates of use. This is the map equation:

\begin{align}\label{map}
L(\mathsf{M}) = q_{\curvearrowright} H(\mathcal{Q}) + \sum_{i=1}^{m}p_{\circlearrowright}^iH(\mathcal{P}^i).
\end{align}

Here $H(\mathcal{Q})$ is the frequency-weighted average length of codewords in the index codebook and $H(\mathcal{P}^i)$ is frequency-weighted average length of codewords in module codebook $i$. Further, the entropy terms are weighted by the rate at which the codebooks are used. With $q_{i \curvearrowright}$ for the probability to exit module $i$, the index codebook is used at a rate $q_{\curvearrowright}=\sum_{i=1}^m q_{i \curvearrowright}$, the probability that the random walker switches modules on any given step. With $p_{\alpha}$ for the probability to visit node $\alpha$, module codebook $i$ is used at a rate $p_{\circlearrowright}^i = \sum_{\alpha \in i} p_\alpha + q_{i \curvearrowright }$, the fraction of time the random walk spends in module $i$ plus the probability that it exits the module and the exit message is used. Now it is straightforward to express the entropies in $q_{i \curvearrowright }$ and $p_\alpha$. For the index codebook, the entropy is

\begin{align}\label{map_master}
H(\mathcal{Q})= -\sum_{i=1}^m\frac{q_{i \curvearrowright}}{\sum_{j=1}^m q_{j \curvearrowright}}\log \left( \frac{q_{i \curvearrowright}}{\sum_{j=1}^m q_{j \curvearrowright}} \right)
\end{align}
and for module codebook $i$ the entropy is
\begin{align}\label{map_module}
{H(\mathcal{P}^i)} &= -{\frac{q_{i \curvearrowright} }{q_{i \curvearrowright }+\sum_{\beta \in i} p_\beta} \log \left( \frac{q_{i \curvearrowright} }{q_{i \curvearrowright }+\sum_{\beta \in i} p_\beta} \right)}\\ \nonumber
&{- \sum_{\alpha \in i}\frac{p_\alpha}{q_{i \curvearrowright }+\sum_{\beta \in i} p_\beta}\log \left( \frac{p_\alpha}{q_{i \curvearrowright }+\sum_{\beta \in i} p_\beta} \right)}.
\end{align}

By combining Eqs.~\ref{map_master} and \ref{map_module} and simplifying, we can write the map equation as:
\begin{align}
L(\mathsf{M}) &= \left(\sum_{i=1}^m q_{i \curvearrowright}\right) \log \left( \sum_{i=1}^m q_{i \curvearrowright} \right)\\ \nonumber
&- 2 \sum_{i=1}^m q_{i \curvearrowright}\log \left( q_{i \curvearrowright} \right) 
 - \sum_{\alpha=1}^{n} p_\alpha \log \left( p_\alpha \right)\label{map5b}\\ \nonumber
&+ \sum_{i=1}^{m}\left(q_{i \curvearrowright }+\sum_{\alpha \in i} p_\alpha \right) \log \left( q_{i \curvearrowright }+\sum_{\alpha \in i} p_\alpha \right).
\end{align}

In this expanded form of the map equation, we note that the term $\sum_{1}^{n} p_\alpha \log \left( p_\alpha \right)$ is independent of partitioning, and elsewhere in the expression $p_\alpha$ appears only when summed over all nodes in a module. Consequently, when we optimize the network partition, it is sufficient to keep track of changes in $q_{i\curvearrowright}$, the rate at which a random walker enters and exits each module, and $\sum_{\alpha \in i} p_\alpha$, the fraction of time a random walker spends in each module. They can easily be derived for any partition of the network, and updating them is a straightforward and fast operation. Any numerical search algorithm developed to find a network partition that optimizes an objective function can be modified to minimize the map equation.

\begin{figure*}[tbp]
\centering
\begin{tabular}{@{\extracolsep{\fill}}l@{\hspace{1em}}l@{\hspace{4em}}l@{\hspace{1em}}l}
	(a) & (b) & (c) & (d) \\
	\includegraphics[width=0.2\textwidth]{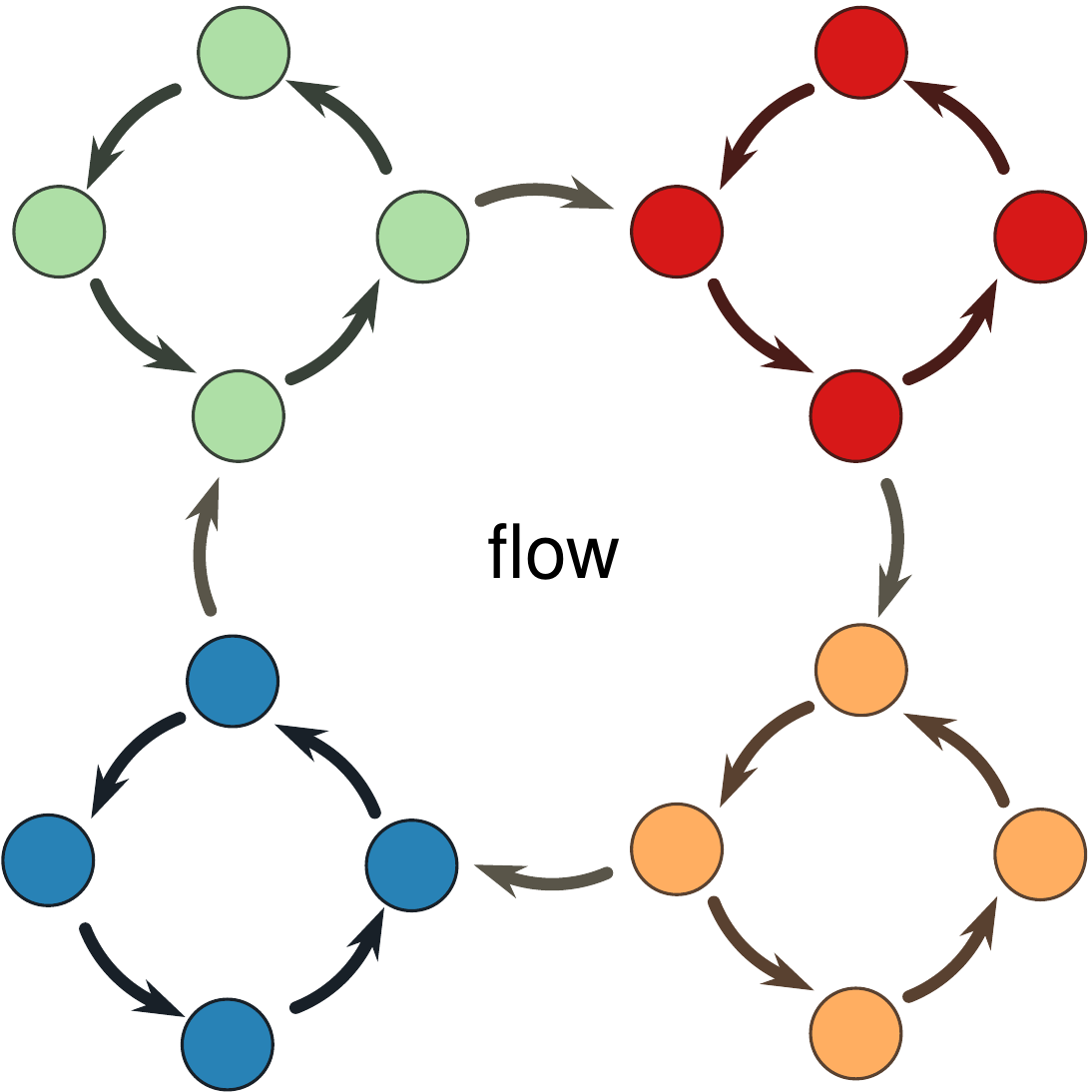} &
	\includegraphics[width=0.2\textwidth]{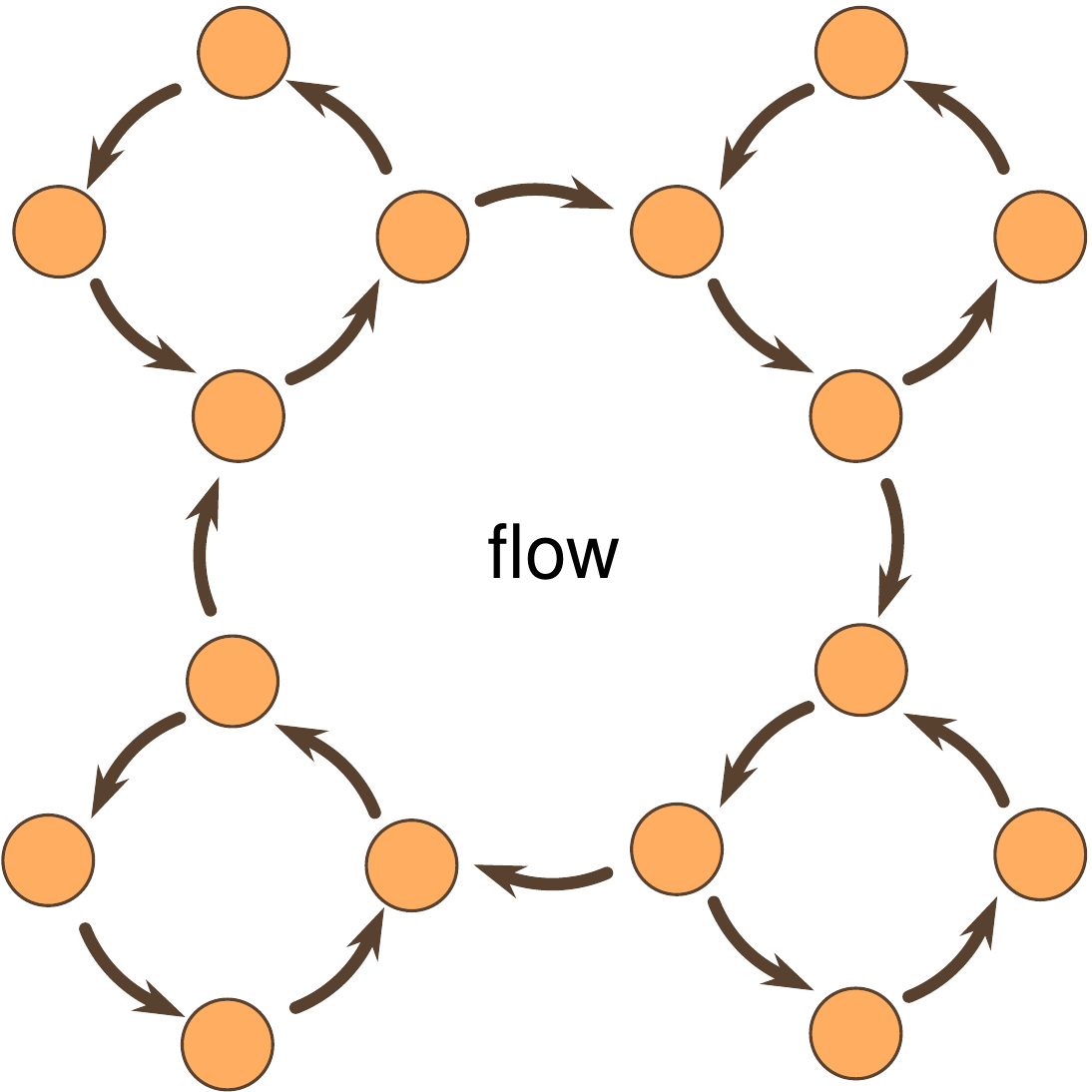} &
	\includegraphics[width=0.2\textwidth]{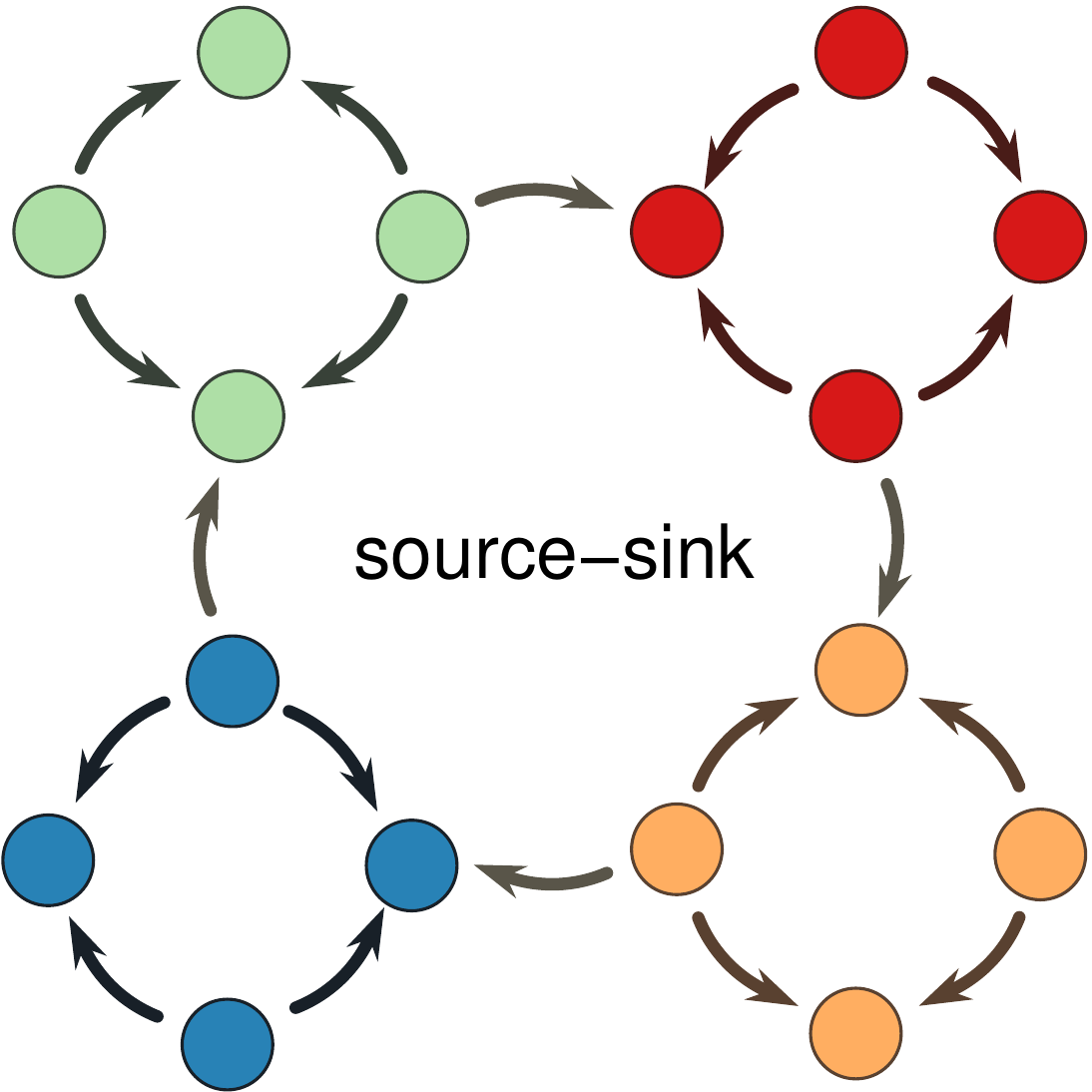}  & 
	\includegraphics[width=0.2\textwidth]{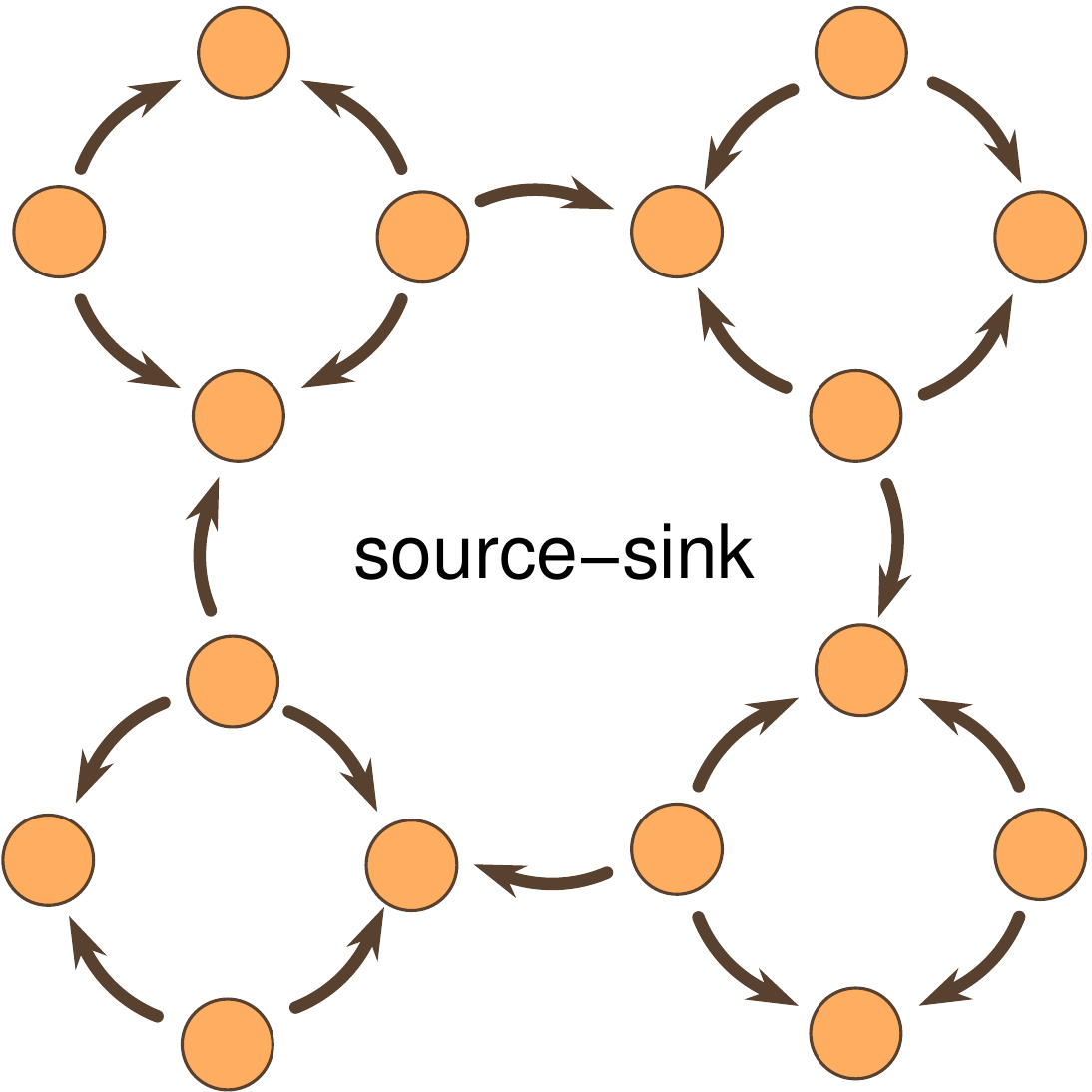}\vspace{3ex} \\ 
	Map equation $L=\mathbf{3.33}$ bits & Map equation $L=3.94$ bits & Map equation $L=4.58$ bits & Map equation $L=\mathbf{3.93}$ bits\\
	Modularity\hspace{1em} $Q = \mathbf{0.55}$ & Modularity\hspace{1em} $Q = 0.00$ & Modularity\hspace{1em} $Q = \mathbf{0.55}$ & Modularity\hspace{1em} $Q = 0.00$
\end{tabular}
\caption{\label{compare}Comparing the map equation with modularity for networks with and without flow. The two sample networks, labelled flow and source-sink, are identical except for the direction of two links in each set of four nodes. The coloring of nodes illustrates alternative partitions. The optimal solutions for the map equation (minimum $L$) and the  modularity (maximum $Q$) are highlighted with boldfaced scores. The directed links in the network in (a) and (b) conduct a system-wide flow with relatively long persistence times in the modules shown in (a). Therefore, the four-module partition in (a) minimizes the map equation. Modularity, because it looks at patterns with high link-weight within modules, also prefers the four-module solution in (a) over the unpartitioned network in (b). The directed links in the network in (c) and (d) represent, not movements between nodes, but rather pairwise interactions, and the source-sink structure induces no flow. With no flow and no regions with long persistence times, there is no use for multiple codebooks and the unpartitioned network in (d) optimizes the map equation. But because modularity only counts weights of links and in-degree and out-degree in the modules, it sees no difference between the solutions in (a)/(c) and (b)/(d), respectively, and again the four-module solution maximizes modularity.}
\end{figure*}

\subsection{Undirected weighted networks}

For undirected networks, the node visit frequency of node $\alpha$ simply corresponds to the relative weight $w_\alpha$ of the links connected to the node. The relative weight is the total weight of the links connected to the node divided by twice the total weight of all links in the network, which corresponds to the total weight of all link-ends. With $w_\alpha$ for the relative weight of node $\alpha$, $w_i=\sum_{\alpha \in i} w_\alpha$ for the relative weight of module $i$, $w_{i\curvearrowright}$ for the relative weight of links exiting module $i$, and $w_\curvearrowright=\sum_{i=1}^{m} w_{i\curvearrowright}$ for the total relative weight of links between modules, the map equation takes the form

\begin{align}
&L(\mathsf{M}) = w_\curvearrowright \log \left( w_\curvearrowright \right) - 2 \sum_{i=1}^m w_{i\curvearrowright}\log \left( w_{i\curvearrowright} \right)\\ \nonumber
&- \sum_{\alpha=1}^{n} w_\alpha \log \left( w_\alpha \right) + \sum_{i=1}^{m}\left(w_{i\curvearrowright} + w_i\right) \log \left(w_{i\curvearrowright} + w_i\right).
\end{align}

\subsection{Directed weighted networks}

For directed weighted networks, we use the power iteration method to calculate the steady state visit frequency for each node. To guarantee a unique steady state distribution for directed networks, we introduce a small teleportation probability $\tau$ in the random walk that links every node to every other node with positive probability and thereby converts the random walker into a \emph{random surfer}. The movement of the random surfer can now be described by an irreducible and aperiodic Markov chain that has a unique steady state by the Perron-Frobineous theorem. As in Google's Page\-Rank algorithm \cite{google}, we use $\tau=0.15$. The results are relatively robust to this choice, but as $\tau \rightarrow 0$, the stationary frequencies may poorly reflect the important nodes in the network as the random walker can get trapped in small clusters that do not point back into the bulk of the network \cite{boldi09}. The surfer moves as follows: at each time step, with probability $1-\tau$, the random surfer follows one of the outgoing links from the node $\alpha$ that it currently occupies to the neighbor node $\beta$ with probability proportional to the weights of the outgoing links $w_{\alpha\beta}$ from $\alpha$ to $\beta$. It is therefore convenient to set $\sum_\beta w_{\alpha\beta} = 1$. With the remaining probability $\tau$, or with probability $1$ if the node does not have any outlinks, the random surfer ``teleports'' with uniform probability to a random node anywhere in the system. But rather than averaging over a single long random walk to generate the ergodic node visit frequencies, we apply the power iteration method to the probability distribution of the random surfer over the nodes of the network. We start with a probability distribution of $p_\alpha=1/n$ for the random surfer to be at each node $\alpha$ and update the probability distribution iteratively. At each iteration, we distribute a fraction $1-\tau$ of the probability flow of the random surfer at each node $\alpha$ to the neighbors $\beta$ proportional to the weights of the links $w_{\alpha\beta}$ and distribute the remaining probability flow uniformly to all nodes in the network. We iterate until the sum of the absolute differences between successive estimates of $p_\alpha$ is less than $10^{-15}$ and the probability distribution has converged.  

Given the ergodic node visit frequencies $p_\alpha$ for $\alpha=1,\ldots,n$ and an initial partitioning of the network, it is easy to calculate the ergodic module visit frequencies $\sum_{\alpha \in i} p_\alpha$ for module $i$. The exit probability for module $i$, with teleportation taken into account, is then
\begin{align}\label{q_jump}
	q_{i \curvearrowright }=\tau\frac{n-n_i}{n}\sum_{\alpha \in i} p_\alpha + (1-\tau)\sum_{\alpha \in i}\sum_{\beta \notin i} p_\alpha w_{\alpha\beta},
\end{align}
where $n_i$ is the number of nodes in module $i$. This equation follows since every node teleports a fraction $\tau(n-n_i)/n$ and guides a fraction $(1-\tau)\sum_{\beta \notin i} w_{\alpha\beta}$ of its weight $p_\alpha$ to nodes outside of its module $i$.

If the nodes represent objects that are inherently different it can be desirable to nonuniformly teleport to nodes in the network. For example, in journal-to-journal citation networks, journals should receive teleporting random surfers proportional to the number of articles they contain, and, in air traffic networks, airports should receive teleporting random surfers proportional to the number of flights they handle. This nonuniform teleportation nicely corrects for the disproportionate amount of random surfers that small journals or small airports receive if all nodes are teleported to with equal probability. In practice, nonuniform teleportation can be achieved by assigning to each node $\alpha$ a normalized teleportation weight $\tau_\alpha$ such that $\sum_{\alpha}\tau_\alpha=1$. With teleportation flow distributed nonuniformly, the numeric values of the ergodic node visit probabilities $p_\alpha$ will change slightly and the exit probability for module $i$ becomes
\begin{align}
	q_{i \curvearrowright }=\tau(1-\sum_{\alpha \in i}\tau_\alpha)\sum_{\alpha \in i} p_\alpha + (1-\tau)\sum_{\alpha \in i}\sum_{\beta \notin i} p_\alpha w_{\alpha\beta}.
\end{align}
This equation follows since every node now teleports a fraction $\tau(1-\sum_{\alpha \in i}\tau_\alpha)$ of its weight $p_\alpha$ to nodes outside of its module $i$.

\section{The map equation compared with modularity}

Conceptually, detecting communities by mapping flow is a very different approach from inferring module assignments for underlying network models. Whereas the former approach focuses on the interdependence of links and the dynamics on the network once it has been formed, the latter one focuses on pairwise interactions and the formation process itself. Because the map equation and modularity take these two disjoint approaches, it is interesting to see how they differ in practice. To highlight one important difference, we compare how the map equation and the generalized modularity, which makes use of information about the weight and direction of links \cite{newman-fast,arenasdirectedweighted,guimeradirected}, operate on networks with and without flow. 

For weighted and directed networks, the modularity for a given partitioning of the network into $m$ modules is the sum of the total weight of all links in each module minus the expected weight
\begin{equation}\label{modularity}
 Q = \sum_{i=1}^{m}\frac{w_{ii}}{w} - \frac{w_{i}^{\mathrm{in}}w_{i}^{\mathrm{out}}}{w^2}.
\end{equation}
Here $w_{ii}$ is the total weight of links starting and ending in module $i$, $w_{i}^{\mathrm{in}}$ and $w_{i}^{\mathrm{out}}$  the total in- and out-weight of links in module $i$, and $w$ the total weight of all links in the network. To estimate the community structure in a network, Eq.~\ref{modularity} is maximized over all possible assignments of nodes into any number $m$ of modules. 

Figure \ref{compare} shows two different networks, each partitioned in two different ways. Both networks are generated from the same underlying network model in the modularity sense: 20 directed links connect 16 nodes in four modules, with equal total in- and out-weight at each module. The only difference is that we switch the direction of two links in each module. Because the weights $w$, $w_{ii}$, $w_{i}^{\mathrm{in}}$, and $w_{i}^{\mathrm{out}}$ are all the same for the four-module partition of the two different networks in Fig.~\ref{compare}(a) and (c), the modularity takes the same value. That is, from the perspective of modularity, the two different networks and corresponding partitions are identical.

However, from a flow-based perspective, the two networks are completely different. The directed links shown in the network in panel (a) and panel (b) induce a structured pattern of flow with long persistence times in, and limited flow between, the four modules highlighted in panel (a). The map equation picks up on these structural regularities, and thus the description length is shorter for the four-module network partition in panel (a) than for the unpartitioned network in panel (b). By contrast, for the network shown in panels (c) and (d), there is no pattern of extended flow at all. Every node is either a source or a sink, and no movement along the links on the network can exceed more than one step in length. As a result, random teleportation will dominate and any partition into multiple modules will lead to a high flow between the modules. For networks with links that do not induce a pattern of flow, the map equation will always be minimized by one single module. 

The map equation captures small modules with long persistence times, and modularity captures small modules with more than the expected number of link-ends, incoming or outgoing.  This example, and the example with directed and weighted networks in ref.~\cite{RosvallBergstrom08}, reveal the effective difference between them.
Though modularity can be interpreted as a one-step measure of movement on a network \cite{delvenne2008sgc}, this example demonstrates that one-step walks cannot capture flow. 

\section{Fast stochastic and recursive search algorithm}

Any greedy (fast but inaccurate) or Monte Carlo-based (accurate but slow) approach can be used to minimize the map equation. To provide a good balance between the two extremes, we have developed a fast stochastic and recursive search algorithm, implemented it in C++, and made it available online both for directed and undirected weighted networks \cite{mapcode}. As a reference, the new algorithm is as fast as the previous high-speed algorithms (the greedy search presented in the supporting appendix of ref.~\cite{RosvallBergstrom08}), which were based on the method introduced in ref.~\cite{clauset-2004-70} and refined in ref.~\cite{wakita}. At the same time, it is also more accurate than our previous high-accuracy algorithm (a simulated annealing approach) presented in the same supporting appendix.

The core of the algorithm follows closely the method presented in ref.~\cite{blondel2008}: neighboring nodes are joined into modules, which subsequently are joined into supermodules and so on. First, each node is assigned to its own module. Then, in random sequential order, each node is moved to the neighboring module that results in the largest decrease of the map equation. If no move results in a decrease of the map equation, the node stays in its original module. This procedure is repeated, each time in a new random sequential order, until no move generates a decrease of the map equation. Now the network is rebuilt, with the modules of the last level forming the nodes at this level. And exactly as at the previous level, the nodes are joined into modules. This hierarchical rebuilding of the network is repeated until the map equation cannot be reduced further. Except for the random sequence order, this is the algorithm described in ref.~\cite{blondel2008}.

With this algorithm, a fairly good clustering of the network can be found in a very short time. Let us call this the core algorithm and see how it can be improved. The nodes assigned to the same module are forced to move jointly when the network is rebuilt. As a result, what was an optimal move early in the algorithm might have the opposite effect later in the algorithm. Because two or more modules that merge together and form one single module when the network is rebuilt can never be separated again in this algorithm, the accuracy can be improved by breaking the modules of the final state of the core algorithm in either of the two following ways:

\begin{itemize}
\item[] \emph{Submodule movements.} First, each cluster is treated as a network on its own and the main algorithm is applied to this network. This procedure generates one or more submodules for each module. Then all submodules are moved back to their respective modules of the previous step. At this stage, with the same partition as in the previous step but with each submodule being freely movable between the modules, the main algorithm is re-applied.

\item[] \emph{Single-node movements.} First, each node is re-assigned to be the sole member of its own module, in order to allow for single-node movements. Then all nodes are moved back to their respective modules of the previous step. At this stage, with the same partition as in the previous step but with each single node being freely movable between the modules, the main algorithm is re-applied.
\end{itemize}

In practice, we repeat the two extensions to the core algorithm in sequence and as long as the clustering is improved. Moreover, we apply the submodule movements recursively. That is, to find the submodules to be moved, the algorithm first splits the submodules into subsubmodules, subsubsubmodules, and so on until no further splits are possible. Finally, because the algorithm is stochastic and fast, we can restart the algorithm from scratch every time the clustering cannot be improved further and the algorithm stops. The implementation is straightforward and, by repeating the search more than once, 100 times or more if possible, the final partition is less likely to correspond to a local minimum. For each iteration, we record the clustering if the description length is shorter than the previously shortest description length. In practice, for networks with on the order of 10,000 nodes and 1,000,000 directed and weighted links, each iteration takes about 5 seconds on a modern PC.

\section*{CONCLUSION}

In this paper and associated interactive visualization \cite{flashapplet}, we have detailed the mechanics of the map equation for community detection in networks \cite{RosvallBergstrom08}. Our aim has been to differentiate flow-based methods such as spectral methods and the map equation, which focus on system behavior once the network has been formed, from methods based on underlying stochastic models such as mixture models and modularity methods, which focus on the network formation process. By comparing how the map equation and modularity operate on networks with and without flow, we conclude that the two approaches are not only conceptually different, they also highlight different aspects of network structure. Depending on the sorts of questions that one is asking, one approach may be preferable to the other. For example, to analyze how networks are formed and to simplify networks for which links do not represent flows but rather pairwise relationships, modularity \cite{girvan} or other topological methods \cite{newman2007mma,clauset2008hsa,RosvallBergstrom07,hofman2008ban} may be preferred. But if instead one is interested in the dynamics on the network, in how local interactions induce a system-wide flow, in the interdependence across the network, and in how network structure relates to system behavior, then flow-based approaches such as the map equation are preferable.


\end{document}